\documentclass[prl,twocolumn]{revtex4}
   \usepackage[english]{babel}
   \usepackage{epsfig}
   \usepackage{color}

\begin{document}

\title{Inelastic resonant tunneling through single molecules and quantum dots:
\newline
spectrum modification due to nonequilibrium effects}

\author{D.A. Ryndyk\footnote{On leave from the Institute for Physics of Microstructures, RAS,
         Nizhny Novgorod, Russia} and J. Keller}

\affiliation{Institut f\"ur Theoretische Physik, Universit\"at Regensburg, Germany}

\begin{abstract}
Resonant electron transport through a mesoscopic region (quantum dot or single
molecule) with electron-phonon interaction is considered  at finite voltage. In this
case the standard Landauer-B\"uttiker approach cannot be applied. Using the
nonequilibrium Green function method we show that due to a nonequilibrium distribution
function of electrons in the mesoscopic region, the inelastic scattering rate and
spectral function of the dot become functions of the voltage and have to be calculated
self-consistently.
\end{abstract}

\date{\today}
\maketitle

During the past several years nonequilibrium quantum transport in nanostructures and,
in particular, transport through mesoscopic tunneling structures, quantum dots (QD)
and single molecules, is in the focus of both experimental and theoretical
investigations because of possible electronic device applications. In this paper we
reconsider one of the basic problems, namely resonant tunneling through a mesoscopic
region with electron-phonon interaction, placed between macroscopic leads. We want to
show that the nonequilibrium distribution function of electrons in the dot changes
significantly the spectral function of the dot.

\begin{figure}
\begin{center}
\epsfxsize=0.7\hsize \epsfbox{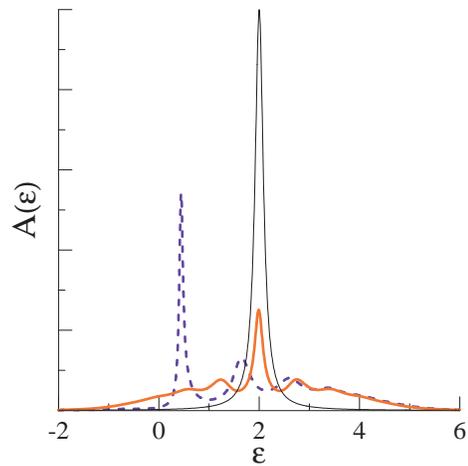}\vskip 0.2cm \caption{(Color online) Spectral function
of the dot with energy level $\epsilon_0=2$. Thin line: no interactions,
dashed line: interaction with phonon, electronic distribution in
equilibrium, thick line: with phonon, including nonequilibrium effects at
high voltage} \label{spectr5}
\end{center}
\end{figure}

\begin{figure}[t]
\begin{center}
\epsfxsize=0.69\hsize \epsfbox{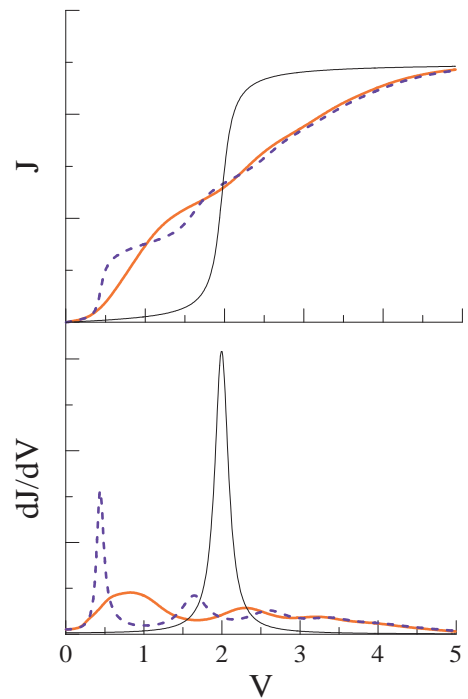} \caption{(Color online) Current-voltage
curves and differential conductance ($\epsilon_0=2$). For the legend of the
curves see Fig. 1} \label{current5}
\end{center}
\end{figure}

In the absence of interactions and in the limit of small voltage the current through a
QD is described by the Landauer-B\"{u}ttiker formula
\begin{equation}\label{LB}
J=\frac{e}{2\pi\hbar}\int T(\epsilon)\left[f^0(\epsilon-e\varphi_L)-
f^0(\epsilon-e\varphi_R)\right]d\epsilon,
\end{equation}
where $T(\epsilon)$ is the transmission coefficient, $\varphi_{L(R)}$ is the
electrical potentials of the left (right) lead. In the case of one resonant level with
energy $\epsilon_0$, the transmission coefficient is given by the Breit-Wigner formula
%
 $ T(\epsilon)={\Gamma^2}/\left[{(\epsilon-\epsilon_0)^2+\Gamma^2}\right]$,
%
where $\Gamma$ is the level width produced by coupling to the leads. The corresponding
spectral function of the dot, current-voltage curve and differential conductance are
shown by the thin lines in Figs.\,\ref{spectr5}, \ref{current5}.

At finite voltage the current can be described by  (\ref{LB}) with a
voltage-independent transmission coefficient only in some special cases, e.g. for
energy-independent coupling to the leads (wide-band limit). In the presence of
interactions (e.g. internal Coulomb interaction and electron-phonon interaction )
$T(\epsilon)$ becomes  temperature and voltage dependent.

The problem of inelastic resonant tunneling of electrons coupled to phonons was first
considered in Refs. \cite{Glazman88jetp,Wingreen88_89,Jonson89prb}. Phonon satellites
are formed in the spectral function and in the differential conductance. Typical
results are shown by dashed lines in Fig.\,\ref{spectr5} and Fig.\,\ref{current5}.
Recently this problem attracted attention after experiments on inelastic electron
transport through single molecules
\cite{Reed97science,Park00nature,Park02nature,Liang02nature,Smith02nature,Zhitenev02prl}.
New theoretical treatments were presented in
Refs.\,\cite{Lundin02prb,Zhu03prb,Braig03prb,Aji03condmat,Mitra03}. In our paper we
study the influence of the nonequilibrium electronic distribution function on  the
phonon self-energy.  We find a significant shift and broadening of the
phonon-satellites in the electronic spectral function. To our knowledge this effect
has not been considered yet.

We use the Nonequilibrium Green Function (NGF) method \cite{Kadanoff62book,Keldysh64},
which now is a standard approach in mesoscopic physics and molecular electronics
\cite{MeirWingreenJauho,Runge92prb,Datta95book,Haug96book}. The advantage of the NGF
formalism is that it can be successfully applied to a variety of systems and problems,
that it is initially exact, and many powerful approximations can be derived from it
(see \cite{Rammer86,Haug96book} and references therein).

We describe the electronic states of the mesoscopic region (QD) by a set a
single-electron states $|\alpha\rangle$ with energies $\epsilon_{\alpha}$. These are
coupled to the free condcuction electrons in the leads by the usual tunneling
Hamiltonian. Furthermore the dot-electrons are  coupled to vibrational modes. We do
not consider Coulomb interaction in the dot, to avoid further complications, such as
Coloumb blockade and Kondo effect. The system is then described by the following model
Hamiltonian:
\begin{equation}\label{H}
  H=H_D + H_L+H_R+H_T+H_{ph},
\end{equation}
with the  dot-Hamiltonian
\begin{equation}\label{H_D}
  H_D=\sum_\alpha(\epsilon_\alpha+e\varphi_D(t))
  d^{\dag}_\alpha
  d_\alpha,
\end{equation}
the Hamiltonian of the right  ($i=R$) and left ($i=L$) lead
\begin{equation}
  H_{i=L,R}=\sum_{k\sigma}(\epsilon_{ik\sigma}+e\varphi_i(t))
  c^{\dag}_{ik\sigma}c_{ik\sigma},
\end{equation}
and the tunneling Hamiltonian
\begin{equation}\label{H_T}
  H_T=\sum_{i=L,R}\sum_{k\sigma,\alpha}\left(V_{ik\sigma,\alpha}
  c^{\dag}_{ik\sigma}d_\alpha+h.c.\right).
\end{equation}
Finally $H_{ph}$ describes phonons and electron-phonon coupling
\begin{equation}\label{H_env}
  H^{ph}_{env}=\sum_q\hbar\omega_qa_q^\dag a_q+
  \sum_\alpha\sum_q\lambda_{q\alpha}(a_{-q}+a_q^\dag)d_\alpha^\dag d_\alpha,
\end{equation}
where $\hbar\omega_q$ is the phonon energy. The quantities $\varphi_i(t)$
and $\varphi_D(t)$ are the electrical potentials of the leads and the dot
The latter will be set to zero in the following.

The current in direction from the left ($i=L$) or right ($i=R$) contact to the dot is
found to be
\begin{equation}\label{J2}
  J_i(t)=\frac{2e}{\hbar}{\rm Re}\left(\sum_{k\sigma,\alpha} V_{ik\sigma,\alpha}
  {G^<_{\alpha,ik\sigma}(t,t)}\right),
\end{equation}
where we define the lead-dot lesser Green function
%
$G^<_{\alpha,ik\sigma}(t_1,t_2)=i\left\langle
c^\dag_{ik\sigma}(t_2)d_\alpha(t_1)\right\rangle$,
%
while the charge of the QD is defined by the dot lesser function
%
 $G^<_{\alpha\beta}(t_1,t_2)=i\left\langle
 d^{\dag}_{\beta}(t_2)d_{\alpha}(t_1)\right\rangle$.

The equation of motion for the lesser Green function $G^<$ contains coupling
to the retarded and advanced Green functions $G^{R,A}$. In order to obtain a
simple Dyson equation it is convenient to combine these functions to a matrix
in Keldysh space \cite{Keldysh64,Rammer86,Haug96book}
\begin{equation}
  \breve{G}=
  \left(\begin{array}{cc} {G}^R
  & G^< \\ 0 & G^A
  \end{array}\right).
\end{equation}
where each component again is a matrix $G_{\eta,\eta'}(t_1,t_2) $ within the
states of the system including all the states of the leads ($\eta =
ik\sigma$) and the states of the dot ($\eta =\alpha$).

For this matrix Green function the nonequilibrium Dyson equation can be written in the
usual way, in differential form
\begin{equation}\label{dyson-keldysh dif}
\left[i\frac{\partial}{\partial t_1}-\breve{H}(t_1)\right]\breve{G}
-\left\{\breve{\Sigma}\breve{G}\right\}=\breve{\delta},
\end{equation}
where $\breve{\delta}=\breve{I}\delta_{\eta\eta'}\delta(t_1-t_2)$, and $\breve{I}$
is the unit matrix in Keldysh space. $\breve{H}(t)=(H_L+H_R+H_D+H_T)\breve{I}$
is the single-particle Hamiltonian.
The curly brackets denote a matrix multiplication and
convolution in time:
$\displaystyle \{AB\}_{\eta\eta'}(t_1,t_2)=\sum_\gamma\int dt_3A_{\eta\gamma}(t_1,t_3)
B_{\gamma\eta'}(t_3,t_2)$.
The self-energy $\breve{\Sigma}$ describes interactions in the dot. The retarded and
advanced Green functions and the spectral function $A=i\left(G^R-G^A\right)$ describe
quasiparticle excitations. The lesser function $G^<$ contains in addition a quantum
distribution function.

In the absence of interactions in the leads (besides the tunneling) one can
derive the following exact expression for the lead-dot function
\cite{MeirWingreenJauho,Haug96book}:
\begin{equation}\label{G_LD}
   \breve{G}_{\alpha,ik\sigma}=\sum_\beta V_{\beta,ik\sigma}\left\{
   \breve{G}_{\alpha\beta}\breve{G}_{ik\sigma}\right\},
\end{equation}
where $\breve{G}_{\alpha\beta}$ is the full dot Green function, while $\breve
{G}_{ik\sigma}$ is the bare Green function of the leads, which is diagonal in
the quantum  numbers $ik\sigma$.

Substituting (\ref{G_LD}) in the general equation (\ref{dyson-keldysh dif})
we obtain the following basic equation for the dot Green function $\breve{\bf
G}(t_1,t_2)\equiv\breve{G}_{\alpha\beta}(t_1,t_2)$:
\begin{equation}\label{DK}
  \left(i\frac{\partial}{\partial t_1}-{\bf H}\right)\breve{\bf G}
  -\left\{\breve{\bf\Sigma}\breve{\bf G}\right\}
 =\breve{\bf I}\delta(t_1-t_2),
\end{equation}
with ${\bf H}\equiv H_{\alpha\beta}=\epsilon_\alpha \delta_{\alpha\beta}$. Here
$\breve{\bf\Sigma}(\epsilon)=
\breve{\bf\Sigma}^{(T)}_{L}(\epsilon)+\breve{\bf\Sigma}^{(T)}_{R}(\epsilon)+
\breve{\bf\Sigma}^{(ph)}(\epsilon)$ is the total self-energy of the dot composed of
the tunneling self-energy
\begin{equation}\label{tse}
   \breve{\bf\Sigma}_{j=L,R}^{(T)}\equiv
  \breve{\Sigma}_{j\alpha\beta}^{(T)}=\sum_{k\sigma}\left\{V^*_{jk\sigma,\alpha}
  \breve{G}_{jk\sigma}V_{jk\sigma,\beta}\right\}.
\end{equation}
and the self-energy $\breve{{\bf\Sigma}}^{(ph)}\equiv
\breve{\Sigma}^{(ph)}_{\alpha\beta}$ describing interactions with phonons.

In the following we consider stationary transport produced by a
time-independent voltage. In that case the Green functions depend only
on the time difference. After Fourier transformation
$\left(G(\epsilon)=\int G(t_1-t_2)e^{i\epsilon(t_1-t_2)}d(t_1-t_2)\right)$ one
obtains
\begin{equation}\label{dk6}
  \left(\epsilon{\bf I}-{\bf H}\right)\breve{\bf G}(\epsilon)
  -\breve{\bf\Sigma}(\epsilon)
  \breve{\bf G}(\epsilon)=\breve{\bf I},
\end{equation}

For the current we obtain the well-known expression \cite{MeirWingreenJauho}
\begin{equation}\label{J}\begin{array}{c}\displaystyle
  J_{i}=\frac{ie}{\hbar}\int\frac{d\epsilon}{2\pi}{\rm Tr}\left\{
  {\bf\Gamma}_i(\epsilon-e\varphi_i)\left({\bf G}^<(\epsilon)+ \right.\right.\\[0.5cm]
  \displaystyle \left.\left.
  +f^0_i(\epsilon-e\varphi_i)
  \left[{\bf G}^R(\epsilon)-{\bf G}^A(\epsilon)\right]\right)\right\},
  \end{array}
\end{equation}
with the level-width function
%
$${\bf\Gamma}_{i=L(R)}(\epsilon)= \Gamma_{i\alpha\beta}(\epsilon) =2\pi\sum_{k\sigma}
V_{ik\sigma,\beta}V^*_{ik\sigma,\alpha}\delta(\epsilon-\epsilon_{ik\sigma}).$$

Equations for the retarded and advanced  functions follow from the diagonal
part of (\ref{dk6})
\begin{equation}\label{GR}
  (\epsilon-\epsilon_\alpha)G^R_{\alpha\beta}-\sum_\gamma
\Sigma^{R}_{\alpha\gamma} G^R_{\gamma\beta}=\delta_{\alpha\beta},
\end{equation}
and the equation for the lesser function (quantum kinetic equation) follows
from the off-diagonal part of (\ref{dk6}) combined with its conjugate
$\breve{\bf G}(\epsilon)\left(\epsilon{\bf I}-{\bf H}\right)
  -\breve{\bf G}(\epsilon)\breve{\bf\Sigma}(\epsilon)=\breve{\bf I}$:
\begin{equation}\label{GK}\begin{array}{c}\displaystyle
  (\epsilon_\beta-\epsilon_\alpha)G^<_{\alpha\beta}- \sum_\gamma
 \left(\Sigma^{R}_{\alpha\gamma}G^<_{\gamma\beta}+\Sigma^{<}_{\alpha\gamma}G^A_{\gamma\beta}
  - \right. \\
\left.-G^{R}_{\alpha\gamma}\Sigma^<_{\gamma\beta}
-G^{<}_{\alpha\gamma}\Sigma^A_{\gamma\beta}   \right) =0. \end{array}
\end{equation}

In the following we will concentrate on the tunneling through one level
$\alpha$ only. Then (\ref{GK}) simplifies further:
\begin{equation}\label{KE}
(\Sigma^R_\alpha-\Sigma^A_\alpha)G^<_\alpha - (G^R_\alpha-G^A_\alpha)
\Sigma^>_\alpha=0.
\end{equation}

For the evaluation of the different terms in this equation we make the
following ansatz for the lesser Green function of the dot
\begin{eqnarray}
&\displaystyle G^<_\alpha(\epsilon) = iA(\epsilon)f(\epsilon), &
\\ &\displaystyle A(\epsilon)=i\left(G^R_\alpha-G^A_ \alpha\right). &
\end{eqnarray}
introducing a general nonequilibrium distribution function $f(\epsilon)$. For the
Green functions of the leads in equilibrium we may use the Fermi function
$f^0(\epsilon)$.

For the retarded tunneling  self-energy $\Sigma^{R(T)}$ one obtains
\begin{equation}\label{SigmaRT}
\Sigma^{R(T)}_{i}=\int\frac{d^3k}{(2\pi)^3}\frac{\vert V_{ik,\alpha}\vert^2}
{\epsilon-\epsilon_k-e\varphi_i+i0}=\Lambda(\epsilon-e\varphi_i)-\frac{i}{2}
\Gamma_{i}(\epsilon-e\varphi_i),
\end{equation}
where $\Lambda$ is the real part of the self-energy, which
usually can be included in the level energy $\epsilon_\alpha$, and $\Gamma$
describes level broadening due to coupling to the leads.
For the corresponding lesser function one finds
\begin{equation}
\Sigma^{<(T)}_i=i\Gamma_i(\epsilon-e\varphi_i) f^0(\epsilon-e\varphi_i).
\end{equation}

In the following we will neglect the energy dependence of the level-width-function
$\Gamma$. Then from (\ref{KE})  we obtain the following kinetic equation for the
nonequilibrium distribution function $f(\epsilon)$:
\begin{equation}\label{f}\begin{array}{c}\displaystyle
\Gamma_L\left[f(\epsilon)-f^0(\epsilon-e\varphi_L)\right]
+\Gamma_R\left[f(\epsilon)-f^0(\epsilon-e\varphi_R)\right]+ \\[0.5cm] \displaystyle
+i\Sigma^{<(ph)}(\epsilon)
+i\left(\Sigma^{R(ph)}(\epsilon)-\Sigma^{A(ph)}(\epsilon)\right)f(\epsilon)=0.
\end{array}
\end{equation}

Finally we have to evaluate the self-energy $\Sigma^{(ph)}$ due to the
interaction with phonons. In order to demonstrate nonequilibrium effects
most clearly we consider only one phonon mode with frequency $\omega_0$.
In the standard self-consistent Born approximation, using Keldysh technique,
one obtains for the  self-energies \cite{Rammer86,Haug96book}
\begin{eqnarray}\label{SigmaRA}
& \displaystyle
  \Sigma^{R(ph)}_\epsilon=\frac{i\lambda^2}{2}\int\frac{d\epsilon'}{2\pi}
  \left(G^{R}_{\epsilon-\epsilon'}D^K_{\epsilon'}+G^K_{\epsilon-\epsilon'}D^{R}_{\epsilon'}\right), & \\
& \displaystyle
  \Sigma^{A(ph)}_\epsilon=\frac{i\lambda^2}{2}\int\frac{d\epsilon'}{2\pi}
  \left(G^{A}_{\epsilon-\epsilon'}D^K_{\epsilon'}+G^K_{\epsilon-\epsilon'}D^{A}_{\epsilon'}\right), &
\end{eqnarray}
\begin{equation}\label{SigmaK}
  \Sigma^{<(ph)}_\epsilon=i\lambda^2\int\frac{d\epsilon'}{2\pi}
  G^<_{\epsilon-\epsilon'}D^<_{\epsilon'}.
\end{equation}
where  $G^K= 2G^< + G^R-G^A$ is the Keldysh Green function.

In the following we do not consider phonon renormalization and polaron effects, and
use bare phonon Green functions:
\begin{eqnarray}
& \displaystyle
  D^R_0(\epsilon)=\frac{1}{\epsilon-\omega_0+i\delta}-\frac{1}{\epsilon+\omega_0+i\delta},
   & \\ & \displaystyle D^<_0(\epsilon)=-2\pi
  i\left[(N_0+1)\delta(\epsilon+\omega_0)+N_0\delta(\epsilon-\omega_0)\right], &
\end{eqnarray}
where $N_0=1/(e^{\omega_0/T}-1)$ is the equilibrium phonon population,
(nonequilibrium effects in the phonon distribution was considered recently in \cite{Mitra03}).
However, we include nonequilibrium effects in the  electron distribution
function while calculating the self-energies.
These self-energies enter the spectral function
$A(\epsilon)=i(G^R_\alpha(\epsilon)-G^A_\alpha(\epsilon))$ of the quantum dot, where
$G^R_\alpha(\epsilon)$ can be found from (\ref{GR})
\begin{equation}
G^R_\alpha(\epsilon)=\frac{1}{\epsilon-\epsilon_\alpha-\Sigma^{R(ph)}_\epsilon+
i(\Gamma_L+\Gamma_R)/2}.
\end{equation}
The real part of the self-energy leads to an energy shift in the spectral function,
which depends on the applied voltage. To take this effect into account properly, we
perform self-consistent calculations of the distribution function $f(\epsilon$)
(\ref{f}) and self-energies (\ref{SigmaRA})-(\ref{SigmaK})  at finite voltage.

\begin{figure}
\begin{center}
\epsfxsize=0.7\hsize \epsfbox{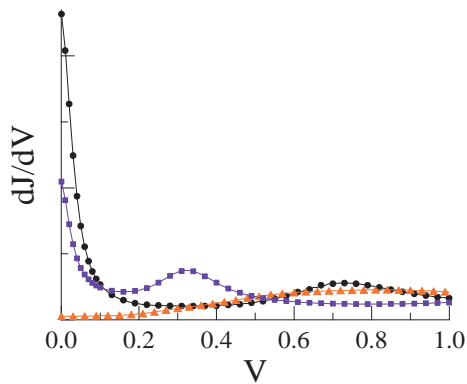} \caption{(Color online) Differential
conductance at small voltages for different positions of the dot energy level
$\epsilon_0=0$ (circles), $\epsilon_0=1$ (squares), and $\epsilon_0=2$
(triangles).} \label{dvcc}
\end{center}
\end{figure}

The spectral function is then used to calculate the current voltage curves. Combining
$J_L$ and $J_R$  (see \cite{MeirWingreenJauho}) the expression for the current can be
written as current
\begin{equation}\label{J_qc}
J=\frac{e}{2\pi\hbar}\frac{\Gamma_L\Gamma_R}{\Gamma_R+\Gamma_L} \int d\epsilon
A(\epsilon)\left[f^0(\epsilon-e\varphi_L)- f^0(\epsilon-e\varphi_R)\right].
\end{equation}
It looks as simple as the Landauer-B\"{u}ttiker formula (\ref{LB}), but it is not
trivial! The spectral density $A(\epsilon)$ now depends on the distribution function
$f(\epsilon)$ and hence the applied voltage, $\varphi_L=-\varphi_R=V/2$. Results for
the current-voltage  curves are presented in Fig.\,\ref{current5} by thick lines.

If the spectral function $A(\epsilon)$ is calculated with help of the equilibrium
distribution function $f^0(\epsilon)$, which does not depend on the voltage, the usual
phonon satellites in the spectral function and differential conductance are obtained
as shown by the dashed lines in Fig.\,\ref{spectr5} and Fig.\,\ref{current5}. But if
one takes into account the nonequilibrium distribution function $f(\epsilon)$, the
spectral function changes with increasing voltage from the equilibrium one to the
function shown as thick line in  Fig.\,\ref{spectr5} for a voltage $V>\epsilon_0$.
Correspondingly, the phonon resonances in the differential conductance are shifted and
suppressed (Fig.\,\ref{current5}, thick line).

Finally in Fig.\,\ref{dvcc} we show the differential conductance for different
positions of the dot level $\epsilon_0$ with respect to the Fermi energy of the leads
(defined in the absence of an applied voltage). For small dot energies the phonon peak
appears at zero voltage and broadening is weak. This property can be used for
experimental investigation of inelastic effects by tuning the dot level with help of a
gate voltage.

In conclusion, we investigated inelastic resonant transport through a mesoscopic
region. As an example, we considered resonant electron tunneling through a single
level coupled to a single phonon mode. This model, although quite simple, is popular
in connection with transport through vibrating molecules. The self-consistent
treatment of the electronic nonequilibrium distribution function shows that the main
peak and phonon side-bands in the differential conductance are shifted considerably
and are essentially broadened as compared with their equilibrium position.

We thank K. Richter and G. Cuniberti for valuable discussions.

\end{document}